\documentclass{Interspeech2024}

\usepackage{setspace} %
\usepackage{booktabs} %
\usepackage{multirow} %
\usepackage[T1]{fontenc}
\usepackage[labelfont=bf]{caption} %
\usepackage{color}
\usepackage[table]{xcolor}
\usepackage{cite}
\usepackage{enumitem}

\usepackage{tikz}
\newcommand{\tikzxmark}{%
\tikz[scale=0.23] {
    \draw[line width=0.7,line cap=round] (0,0) to [bend left=6] (1,1);
    \draw[line width=0.7,line cap=round] (0.2,0.95) to [bend right=3] (0.8,0.05);
}}
\newcommand{\tikzcmark}{%
\tikz[scale=0.23] {
    \draw[line width=0.7,line cap=round] (0.25,0) to [bend left=10] (1,1);
    \draw[line width=0.8,line cap=round] (0,0.35) to [bend right=1] (0.23,0);
}}

\makeatletter
\def\bstctlcite{\@ifnextchar[{\@bstctlcite}{\@bstctlcite[@auxout]}}
\def\@bstctlcite[#1]#2{\@bsphack
  \@for\@citeb:=#2\do{%
    \edef\@citeb{\expandafter\@firstofone\@citeb}%
    \if@filesw\immediate\write\csname #1\endcsname{\string\citation{\@citeb}}\fi}%
  \@esphack}
\makeatother

\interspeechcameraready

\title{Beyond Performance Plateaus: A Comprehensive Study on Scalability in Speech Enhancement}

\name[affiliation={1,2}]{Wangyou}{Zhang}
\name[affiliation={3}]{Kohei}{Saijo}
\name[affiliation={2}]{Jee-weon}{Jung}
\name[affiliation={1,2}]{Chenda}{Li}
\name[affiliation={2}]{Shinji}{Watanabe}
\name[affiliation={1}]{Yanmin}{Qian}

\address{
  $^1$Shanghai Jiao Tong University, China \quad
  $^2$Carnegie Mellon University, USA \\
  $^3$Waseda University, Japan}
\email{wyz-97@sjtu.edu.cn}


\begin{document}
\bstctlcite{IEEEexample:BSTcontrol} %

\maketitle

\begin{abstract}

    Deep learning-based speech enhancement (SE) models have achieved impressive performance in the past decade.
    Numerous advanced architectures have been designed to deliver state-of-the-art performance; however, their scalability potential remains unrevealed.
    Meanwhile, the majority of research focuses on small-sized datasets with restricted diversity, leading to a plateau in performance improvement.
    In this paper, we aim to provide new insights for addressing the above issues by exploring the scalability of SE models in terms of architectures, model sizes, compute budgets, and dataset sizes.
    Our investigation involves several popular SE architectures and speech data from different domains.
    Experiments reveal both similarities and distinctions between the scaling effects in SE and other tasks such as speech recognition.
    These findings further provide insights into the under-explored SE directions, e.g., larger-scale multi-domain corpora and efficiently scalable architectures.
\end{abstract}

\vspace{-8pt}
\section{Introduction}
\label{sec:intro}
\vspace{-4pt}
Speech enhancement (SE) is the task of removing undesired signals from the input speech~\cite{Speech-Loizou2013}, which often includes noise and reverberation.
The last decade has witnessed significant progress in deep learning-based SE approaches in both frequency and time domains.
The state-of-the-art (SOTA) SE approaches often feature a carefully designed architecture with sophisticated interactions between different features.
Typical architectures of SE models include the U-Net structure~\cite{Wave_U_Net-Stoller2018,rouard2023hybrid}, dilated convolution-based structure~\cite{Conv_TasNet-Luo2019}, and dual-path structure with recurrent neural network~(RNN) or attention~\cite{Dual_path-Luo2020,Dual_Path-Chen2020,Attention-Subakan2021,TF_GridNet-Wang2023}.

While a wide variety of SE models have been developed in recent years and demonstrate surprisingly strong performance on specific datasets, there remain several issues.
	1) Most SE research has been conducted on small-scale datasets (e.g., VoiceBank+DEMAND~\cite{Speech-Valentini-Botinhao2016}) with limited diversity. The limited data amount hinders us from understanding the scalability of the SE approach. The limited evaluation condition also makes it difficult to understand the generalizability and robustness in realistic applications.
	2) While it has been commonly observed that more parameters and higher complexity often result in better performance, the scaling effect of model complexity along with data scales has not been well studied in the SE literature. In particular, it is still unclear whether a large amount of data and high-complexity models are the best way to achieve the best performance. %
	3) Compared to the recent trend of large models showing unprecedented performances in automatic speech recognition (ASR) and large language models (LLMs)~\cite{Survey-Zhao2023,Robust-Radford2023,OWSMv3.1-Peng2024,Towards-Wu2024}, this area of research is under-explored in SE.
    This calls for an extensive investigation into the scaling law of SE models as an important step in building a large SE foundation model.

As the first step to address the above issues, in this paper, we aim to explore the scalability of single-channel SE models.
Following existing studies on scaling laws in ASR~\cite{Scaling-Droppo2021,Scaling-Gu2023}, natural language processing~\cite{Scaling-Kaplan2020}, and computer vision~\cite{Scaling-Zhai2022}, we explore several typical scaling factors.
Specifically, they include model architectures, model sizes, compute budgets, and dataset sizes.
We further take the model setup (causal or non-causal) into account to encompass both real-time and offline applications.

While our investigation is driven by motivations akin to~\cite{Complexity-Chen2024}, it distinguishes itself through several novel findings and complements the existing work in five aspects: 1) larger model complexity; 2) extensive investigation of both causal and non-causal setups; 3) multiple data scales; 4) extended coverage of SE sub-tasks; and 5) comprehensive multi-domain evaluation.
Our key contributions are summarized below:
\begin{enumerate}[wide,label={\arabic*)},labelindent=6pt]
	\item By covering a wider range of model complexity, we reveal the \emph{distinct potentials of different SE architectures in low and high complexity regions}.
	Among them, BSRNN and TF-GridNet show exceptional scalabilities respectively in low and (relatively) high complexity ranges.
	\item We show that \emph{scaling model complexity along with multi-domain data sizes consistently improves all metrics}.
	This aligns with the findings in large-scale ASR~\cite{Robust-Radford2023,OWSMv3.1-Peng2024} and LLMs~\cite{Survey-Zhao2023}.
	\item SE models, compared to ASR and LLMs, \emph{particularly suffer from data mismatch} due to the limitations of simulation-based training data.
	As a result, largely increasing simulated data based on the same source corpora is not as effective as increasing data diversity.
	Although we successfully scaled the data up to 157 h, further scaling up encountered various issues such as scarcity of high quality data, limited diversity, and highly imbalanced data distribution.
	This calls for the construction of larger-scale multi-domain SE corpora, as we are still far away from the data scale in ASR research (e.g., 1M h).
	\item Although RNN-based architectures~\cite{Efficient-Yu2023,TF_GridNet-Wang2023} have excelled in popular SE benchmarks, they are highly inefficient in scaling up, demanding much larger compute due to limited parallelization.
	\emph{We are still lacking an SE counterpart to transformers in ASR and LLMs that have well-recognized scalability and outstanding performance.}
	\item Finally, we show that model parameters are not very informative when comparing different architectures, as they usually have distinct parameter efficiency.
	Instead, \emph{model complexity in terms of computational costs (\#MACs) should be compared}.
\end{enumerate}

\vspace{-6pt}
\section{Model and experimental design}
\label{sec:design}
\vspace{-3pt}

\subsection{Models}
\label{ssec:model}
\vspace{-3pt}
We aim to investigate SOTA SE models with diverse architectures to gain insights into their potential scalability.
Specifically, we experiment with band-split RNN (BSRNN)~\cite{Efficient-Yu2023,High-Yu2023}, Conv-TasNet~\cite{Conv_TasNet-Luo2019}, DEMUCS-v4~\cite{rouard2023hybrid}, and TF-GridNet~\cite{TF_GridNet-Wang2023}.
We briefly introduce each model below, while their model configurations are presented in Tables~\ref{tab:models} and \ref{tab:models_other}.

\begin{table}[t]
    \setstretch{0.9}
    \caption{Configurations of BSRNN models\protect\footnotemark.}
    \label{tab:models}
    \centering
    \setlength{\tabcolsep}{3pt}
    \begin{tabular}{lcccc}
        \toprule
        \multirow{2}{*}{\textbf{Model}} & \multirow{2}{*}{\textbf{Causal}} & \multirow{2}{*}{\textbf{\#Params (M)}} & \multicolumn{2}{c}{\textbf{\#MACs (G/s)}} \\
        & & & \textbf{16 kHz} & \textbf{48 kHz} \\
        \midrule
        \multicolumn{2}{l}{\cellcolor[HTML]{EEEEEE}\emph{BSRNN}} & \multicolumn{3}{r}{\cellcolor[HTML]{EEEEEE}(sampling-frequency-independent)} \\
        \multirow{2}{*}{xtiny} & \tikzcmark & 0.5 & 0.1 & 0.4 \\
        & \tikzxmark & 0.5 & 0.2 & 0.6 \\
        \multirow{2}{*}{tiny} & \tikzcmark & 1.3 & 0.6 & 1.7 \\
        & \tikzxmark & 1.5 & 0.7 & 2.2 \\
        \multirow{2}{*}{small} & \tikzcmark & 4.1 & 2.1 & 6.4 \\
        & \tikzxmark & 4.8 & 2.8 & 8.5 \\
        \multirow{2}{*}{medium} & \tikzcmark & 14.3 & 8.4 & 25.2 \\
        & \tikzxmark & 16.9 & 11.2 & 33.4 \\
        \multirow{2}{*}{large} & \tikzcmark & 52.9 & 33.4 & 99.9 \\
        & \tikzxmark & 63.1 & 44.3 & 132.5 \\
        \multirow{2}{*}{xlarge} & \tikzcmark & 83.6 & 66.1 & 197.7 \\
        & \tikzxmark & 104.1 & 87.9 & 262.3 \\
        \bottomrule
    \end{tabular}%
\end{table}

\emph{BSRNN}~\cite{Efficient-Yu2023} is a dual-path T-F domain SE model that can handle different sampling frequencies (SF).
It reduces the frequency dimension with a hand-crafted band division to lower the complexity.
Each T-F bin is projected to a high-dimensional embedding for RNN-based dual-path modeling.
We adopt the improved design in~\cite{High-Yu2023} which combines masking and mapping to obtain the enhanced spectrum.
In~\cite{Complexity-Chen2024}, BSRNN has shown superior performance over other architectures across different complexity.
Therefore, we first conduct extensive experiments with this model to investigate the scaling effect with different model setups (causal or non-causal) as shown in Table~\ref{tab:models}.

\emph{Conv-TasNet}~\cite{Conv_TasNet-Luo2019} is a time-domain SE model, which consists of convolution-based encoder/decoder and stacked temporal convolutional networks (TCN). It features small kernel/stride sizes in the learnable encoder/decoder and a large receptive field thanks to the stacked TCNs.

\emph{DEMUCS-v4}~\cite{rouard2023hybrid} is a hybrid time and time-frequency (T-F) domain SE model with a U-Net architecture.
It consists of two parallel branches, i.e., time-domain U-Net and T-F domain U-Net.
The bottleneck features from both branches are fused via cross-attention in a cross-domain transformer encoder, and both branches' outputs are summed to obtain the enhanced speech.

\footnotetext{\label{footnote1}Detailed hyperparameter configurations of all models are available at \url{https://github.com/Emrys365/se-scaling}.}

\emph{TF-GridNet}~\cite{TF_GridNet-Wang2023} is the SOTA single-channel SE approach based on T-F dual-path modeling.
It also leverages RNNs for dual-path modeling and further enhances the sequence modeling capability by utilizing adjacent frames/frequency bins.
In addition, a cross-frame self-attention module is inserted after each dual-path modeling block to better exploit the global information.
In our preliminary experiments, we noticed that the original TF-GridNet already has very high computational costs, making it expensive to scale up.
Therefore, we focus on smaller-sized configurations for this model as shown in Table~\ref{tab:models_other}.

\begin{table}[t]
    \setstretch{0.9}
    \caption{Configurations of other SE models\protect\footref{footnote1}.}
    \label{tab:models_other}
    \centering
    \setlength{\tabcolsep}{3pt}
    \begin{tabular}{lcccc}
        \toprule
        \multirow{2}{*}{\textbf{Model}} & \multirow{2}{*}{\textbf{Causal}} & \multirow{2}{*}{\textbf{\#Params (M)}} & \multicolumn{2}{c}{\textbf{\#MACs (G/s)}} \\
        & & & \textbf{16 kHz} & \textbf{48 kHz} \\
        \midrule
        \multicolumn{2}{l}{\cellcolor[HTML]{EEEEEE}\emph{Conv-TasNet}} & \multicolumn{3}{r}{\cellcolor[HTML]{EEEEEE}(input is always resampled to 48 kHz)} \\
        small & \tikzxmark & 1.1 & - & 8.9 \\
        medium & \tikzxmark & 14.3 & - & 18.7 \\
        large & \tikzxmark & 52.6 & - & 47.2 \\
        xlarge & \tikzxmark & 103.9 & - & 85.4 \\
        \hline
        \multicolumn{2}{l}{\cellcolor[HTML]{EEEEEE}\emph{DEMUCS-v4}} & \multicolumn{3}{r}{\cellcolor[HTML]{EEEEEE}(input is always resampld to 48 kHz)} \\
        tiny & \tikzxmark & 1.0 & - & 1.0 \\
        small & \tikzxmark & 4.1 & - & 3.5 \\
        medium & \tikzxmark & 16.2 & - & 13.0 \\
        large & \tikzxmark & 26.9 & - & 17.2  \\
        xlarge & \tikzxmark & 79.3 & - & 40.7 \\
        \hline
        \multicolumn{2}{l}{\cellcolor[HTML]{EEEEEE}\emph{TF-GridNet}} & \multicolumn{3}{r}{\cellcolor[HTML]{EEEEEE}(sampling-frequency-independent)} \\
        xxtiny & \tikzxmark & 0.1 &  1.9 & 5.6 \\
        xtiny & \tikzxmark & 0.5 &  7.4 & 21.7 \\
        tiny & \tikzxmark & 1.5 & 24.1 & 70.5 \\
        small & \tikzxmark & 5.7 & 89.5 & 261.8 \\
        \bottomrule
    \end{tabular}%
\end{table}

To allow the processing of different SFs, the simple solution is to always resample the input signal to 48 kHz for all SE models regardless of their original SF $f_{\text{org}}$.
The output is then downsampled back to $f_{\text{org}}$ for generating the enhanced output as well as for loss computation.
But specifically for T-F dual-path models (BSRNN and TF-GridNet), we adopt a different approach by applying the sampling-frequency-independent (SFI) design as proposed in~\cite{Toward-Zhang2023}.
It uses adaptive short-time Fourier transform (STFT) window/hop sizes to handle different SFs without the need for resampling.
During training, we augment each training sample by randomly downsampling it to a new SF $f_{\text{org}}^{\text{aug}}$ in $\{8, 16, 24, 32, 44.1, 48\}$ kHz.
It assists the trained SE models in generalizing to different SFs.
The SF flows of both non-SFI and SFI models are summarized below:
{\setlength{\abovedisplayskip}{3pt}
\setlength{\belowdisplayskip}{3pt}
\begin{align}
&\resizebox{0.9\columnwidth}{!}{%
$
    f_{\text{org}} \xrightarrow[]{\text{augment}} f_{\text{org}}^{\text{aug}} \xrightarrow[]{\text{upsample}} 48 \text{\ kHz} \xrightarrow[\text{model}]{\text{non-SFI}} 48 \text{kHz} \xrightarrow[]{\text{downsample}} f_{\text{org}}^{\text{aug}} \,,
$
}%
\label{eq:non_sfi}\\
&\resizebox{0.4\columnwidth}{!}{%
$
    f_{\text{org}} \xrightarrow[]{\text{augment}} f_{\text{org}}^{\text{aug}} \xrightarrow[\text{model}]{\text{\ SFI}} f_{\text{org}}^{\text{aug}} \,.
$
}%
\label{eq:sfi}
\end{align}
}

The L1-based time-domain plus frequency-domain multi-resolution loss~\cite{Towards-Lu2022} is adopted to train all SE models.
We use four STFT window sizes $\{256, 512, 768, 1024\}$ when calculating the multi-resolution loss.
We use the ESPnet-SE toolkit~\cite{ESPnet_SE-Li2021,ESPnet_SE-Lu2022} for all experiments.
All models are trained until convergence (up to 3M iterations)\footref{footnote1}.

\begin{figure*}
  \centering
  \includegraphics[width=\textwidth]{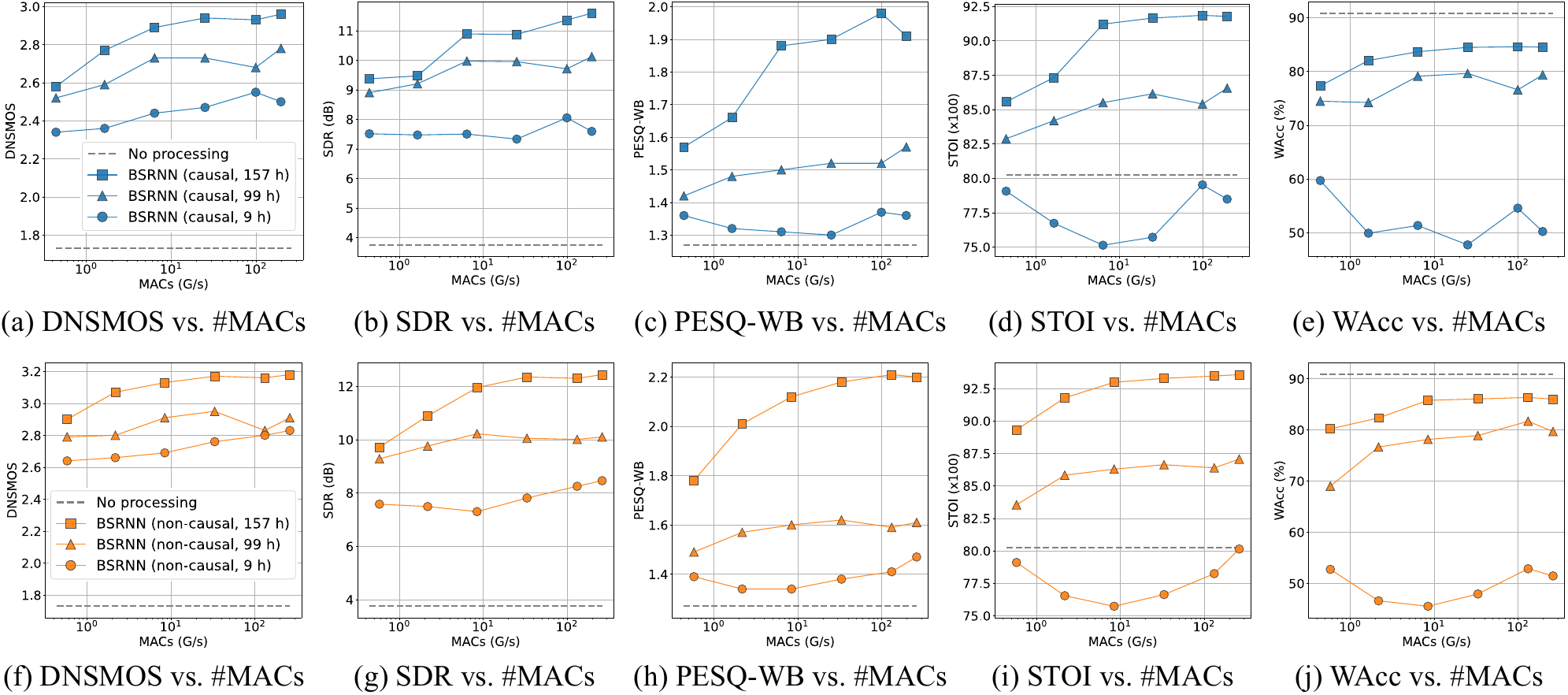}
  \caption{Scaling effect of BSRNN with respect to model complexity (\#MACs at 48 kHz). Each data point corresponds to an independent model. Causal model setups: (a)--(e). Non-causal model setups: (f)--(j).}
  \label{fig:bsrnn_scaling}
  \vspace{-14pt}
\end{figure*}
\vspace{-6pt}
\subsection{Data details}
\label{ssec:data}
\vspace{-4pt}
As shown in Table~\ref{tab:data}, we combine three datasets from different domains to train all aforementioned SE models, i.e., VoiceBank+DEMAND~\cite{Speech-Valentini-Botinhao2016}, DNS-2020 challenge data~\cite{INTERSPEECH2020-Reddy2020}, and WHAMR!~\cite{WHAMR-Maciejewski2019}.
They cover various conditions including noise, reverberation, and different SFs.
We investigate three different data scales \{8.8, 98.8, 156.8\} h by combining the listed datasets. %

For evaluation, we combine five different test sets as listed in Table~\ref{tab:data_eval}.
The first three correspond to the test sets of the corpora mentioned above, representing the matched conditions.
The rest two are challenge datasets from CHiME-4~\cite{Analysis-Vincent2017} and REVERB~\cite{REVERB-Kinoshita2013}, which provide both simulation and real-recorded data for evaluation.
They represent mismatched conditions as they contain unseen speech/noise/reverberation during training.

\vspace{-4pt}
\subsection{Evaluation metrics}
\label{ssec:metric}
We adopt the following evaluation metrics to analyze the scaling effect from multiple views: PESQ-WB~\cite{PESQ-Rix2001}, short-time objective intelligibility (STOI)~\cite{STOI-Taal2011}, signal-to-distortion ratio (SDR)~\cite{SDR-Vincent2006}, DNSMOS~\cite{DNSMOS-Reddy2022} and word accuracy (WAcc)\footnote{WAcc is equal to $1 -$ word error rate (WER).}.
The first three metrics are intrusive objective measures that require well-aligned reference signals.
Therefore, they are only calculated on simulated test samples.
The DNSMOS OVRL score is a non-intrusive metric predicted by a pre-trained neural network that does not require the reference signal as input.
The WAcc is evaluated using the open-source large-scale pre-trained ASR model --- OWSM v3.1~\cite{OWSMv3.1-Peng2024}.
For all metrics, a higher value indicates better performance.

\begin{table}[t]
    \setstretch{0.9}
    \caption{Details of training data.}
    \label{tab:data}
    \centering
    \setlength{\tabcolsep}{3pt}
    \resizebox{\columnwidth}{!}{%
    \begin{tabular}{llccc}
        \toprule
        \multirow{2}{*}{\textbf{No.}} & \multirow{2}{*}{\textbf{Dataset}} & \multirow{2}{*}{\textbf{Condition}} & \multicolumn{2}{c}{\textbf{Duration}} \\
        & & & \textbf{training} & \textbf{validation} \\
        \midrule
        1 & VCTK+DEMAND~\cite{Speech-Valentini-Botinhao2016} & Noisy, 48 kHz & 8.8 h & 0.6 h \\
        2 & DNS-2020~\cite{INTERSPEECH2020-Reddy2020} & Noisy, 16 kHz & 90.0 h & 10.0 h \\
        3 & WHAMR!~\cite{WHAMR-Maciejewski2019} & Noisy, reverberant, 16 kHz & 58.0 h & 14.7 h \\
        \midrule
        \multicolumn{2}{l}{\cellcolor[HTML]{EEEEEE}\textbf{Scale 1}: No.~1 (\textasciitilde{}9 h)} & \multicolumn{3}{l}{\cellcolor[HTML]{EEEEEE}\textbf{Scale 2}: Nos.~1 + 2 (\textasciitilde{}99 h)} \\
        \multicolumn{5}{l}{\cellcolor[HTML]{EEEEEE}\textbf{Scale 3}: Nos.~1 + 2 + 3 (\textasciitilde{}157 h)} \\
        \bottomrule
    \end{tabular}%
    }
\end{table}
\begin{table}[t]
    \setstretch{0.9}
    \caption{Details of combined evaluation data.}
    \label{tab:data_eval}
    \centering
    \setlength{\tabcolsep}{3pt}
    \resizebox{\columnwidth}{!}{%
    \begin{tabular}{llcc}
        \toprule
        \textbf{No.} & \textbf{Dataset} & \textbf{Condition} & \textbf{Duration} \\
        \midrule
        1 & VCTK+DEMAND~\cite{Speech-Valentini-Botinhao2016} & Noisy, 48 kHz & 0.6 h \\
        2 & DNS-2020~\cite{INTERSPEECH2020-Reddy2020} & Noisy, 16 kHz & 0.4 h (no reverb) \\
        3 & WHAMR!~\cite{WHAMR-Maciejewski2019} & Noisy, reverberant, 16 kHz & 9.0 h \\
        4 & CHiME-4~\cite{Analysis-Vincent2017} & Noisy, 16 kHz & 2.3 h (Simu) + 2.2 h (Real) \\
        5 & REVERB~\cite{REVERB-Kinoshita2013} & Reverberant, 16 kHz & 4.8 h (Simu) + 0.7 h (Real) \\
        \bottomrule
    \end{tabular}%
    }
\end{table}

\setcounter{figure}{2}
\begin{figure*}
  \centering
  \includegraphics[width=0.85\textwidth]{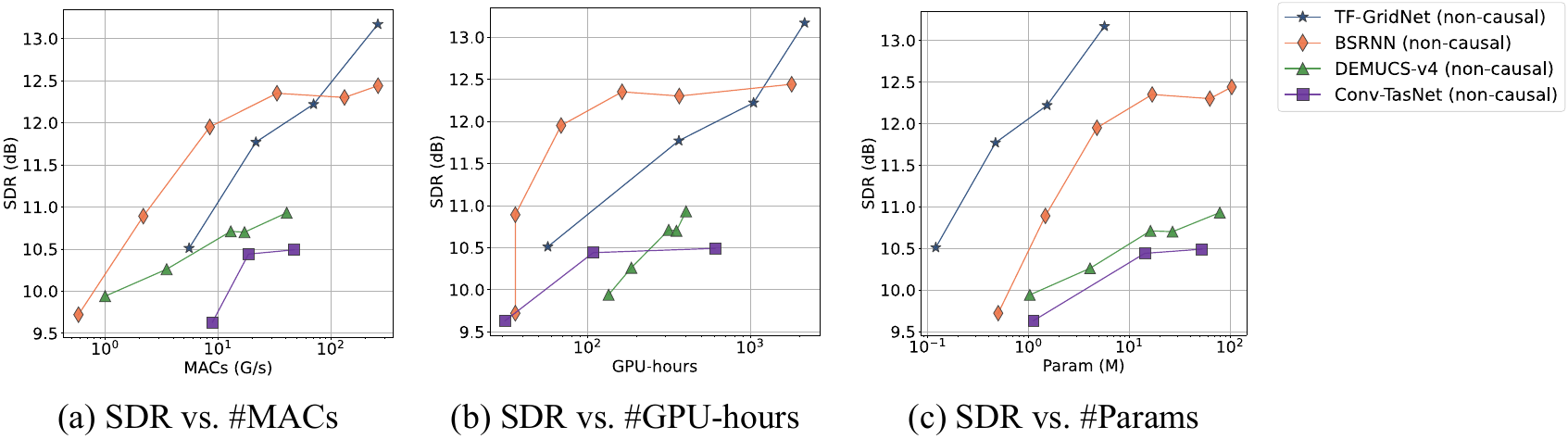}
  \caption{Scaling effect of non-causal SE models with different architectures on 157 h of training data.}
  \label{fig:scaling}
  \vspace{-10pt}
\end{figure*}
\setcounter{figure}{1}
\begin{figure}[t]
  \centering
  \includegraphics[width=0.94\columnwidth]{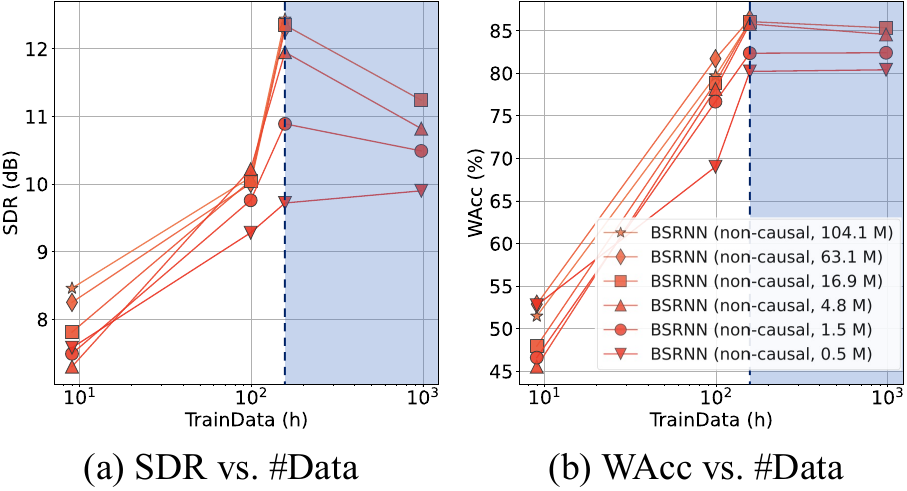}
  \caption{Scaling effect of non-causal BSRNN with respect to dataset sizes (\#Data). Main results are shown in white regions, while shaded regions illustrate the degradation caused by improper data scaling (detailed in \S~\ref{ssec:exp_causality}).}
  \label{fig:bsrnn_scaling2}
\end{figure}
\setcounter{figure}{3}
\vspace{-6pt}
\section{Results}
\label{sec:exp}
\vspace{-4pt}
\subsection{Causal vs. non-causal: a case study on BSRNN}
\label{ssec:exp_causality}

We first study the scaling effect of non-causal and causal models based on BSRNN, as it has shown superior performance in~\cite{Complexity-Chen2024}.
We train BSRNN models with different configurations as listed in Tables~\ref{tab:models} and \ref{tab:data}, and then evaluate them on all test data shown in Table~\ref{tab:data_eval}.
All results presented below represent averages across all test corpora, unless specified otherwise.

Fig.~\ref{fig:bsrnn_scaling} shows the scaling effect of causal (above) and non-causal (below) models with respect to model complexity which is represented by \#MACs at 48 kHz.
It is evident that both causal and non-causal models exhibit similar trends across different metrics. In particular, the common observations include:
\begin{itemize}
	\item \emph{Increasing the model complexity generally improves all metrics when sufficient training data are available.}
	This aligns with the observation in~\cite{Complexity-Chen2024}.
	One major exception is the SDR, PESQ-WB, and STOI metrics on the smallest data scale (9 h), which degrade initially and then start to improve.
	The initial degradation can be attributed to overfitting due to limited training data.
	Combined with the latter, this phenomenon is known as the ``double descent''~\cite{Reconciling-Belkin2019, Deep-Nakkiran2021}, which shows that modern deep architectures can recover from overfitting by further increasing model parameters.
	However, it has been rarely reported in the area of speech processing.
	In our context, this observation implies that the BSRNN with model complexity higher than 10 G/s has more than enough capacities of modeling the 9 h training data.
	This might also apply to other SE architectures.
	These results may suggest that SE research should focus more on larger corpora to avoid overkill on small datasets such as VCTK+DEMAND~\cite{Speech-Valentini-Botinhao2016} with high-complexity SE models.

	\item \emph{With more training data, the model performance tends to benefit more from increased model complexity.} The models trained on 157 h data gain more performance improvement than other models.
	\item \emph{When the model complexity reaches a certain threshold, the initially linear scaling relationship between performance and log-scale \#MACs tend to converge.} The initially linear scaling part coincides with the observation in~\cite{Complexity-Chen2024}, which only investigates the model complexity between 50 M/s and 15 G/s. Our study reveals that this scaling effect does not hold when we further increase the model complexity, implying that more training data is needed when further scaling up the model.
    \item \emph{More complex models tend to achieve better downstream ASR performance (WAcc), but they still underperform the baseline without SE (no processing).} Similar issues are also reported in prior works~\cite{Human-Eskimez2021,How-Iwamoto2022}, where the degradation is attributed to the artifacts introduced by SE models on mismatched data.
\end{itemize}

Comparing causal and non-causal models in Fig.~\ref{fig:bsrnn_scaling}, we can also see that non-causal models consistently outperform causal models in all metrics.
The performance difference is especially enlarged when more training data (e.g., 157 h) are available, indicating a much higher model capacity in non-causal models.

Furthermore, we analyze the scaling effect of model performance with respect to dataset sizes.
In Fig.~\ref{fig:bsrnn_scaling2}, we only present the results of non-causal models in terms of SDR and WAcc metrics, and omit the other results as they share a similar trend.
We can see that \emph{all metrics keep improving monotonically with more training data}, which is in line with common observations.
Moreover, larger models tend to benefit more from increased data, indicating that a higher model capacity allows for better utilization of the data.
\emph{Note that this does not imply that simply increasing data is always helpful.}
In fact, when we attempted to further increase the dataset size by simply simulating more DNS-2020 data on top of ``scale 3'' in Table~\ref{tab:data} (from 157 h to \textasciitilde{}900h), the overall performance was even degraded.
This is shown in the shaded regions in Fig.~\ref{fig:bsrnn_scaling2}.
The major cause roots in the highly imbalanced data distribution, with 93\% DNS-2020, 1\% VCTK+DEMAND, and 6\% WHAMR! data.
The trained SE model is thus biased towards DNS-2020 data and suffers from mismatch during evaluation.
A similar observation has also been reported in~\cite{Scaling-Isik2024}, where increasing mismatched pre-training data does not always improve the downstream finetuning performance.
This implies the importance of well-balanced and multi-domain data when scaling up the training dataset, which has often been overlooked in prior SE studies.

\vspace{-6pt}
\subsection{Comparison of different SE architectures}
\label{ssec:exp_arch}
\vspace{-3pt}
Next, we investigate the scaling effect of different SE architectures.
Given the findings from BSRNN in the previous section,  we mainly focus on the 157 h data scale and non-causal model setup for the best performance.
In Fig.~\ref{fig:scaling}, we explore three different scaling factors, i.e., model complexity (\#MACs at 48 kHz), model parameters (\#Params), and compute budget (\#GPU-hours).

In Fig.~\ref{fig:scaling} (a), we observe that BSRNN outperforms others in the low model complexity region ($<$100 G/s), while TF-GridNet dominates the higher model complexity region.
They are both T-F domain dual-path models, the recently developed SOTA architecture.
The U-Net-based DEMUCS-v4 models performs better than TCN-based Conv-TasNet, but still lags behind BSRNN.
This implies that T-F domain dual-path models have great potential in both low and high-complexity configurations, where BSRNN and TF-GridNet can be chosen depending on the complexity budget.
However, \emph{no single architecture in our investigation can dominate all complexity conditions}.

In Fig.~\ref{fig:scaling} (b), we present the scaling effect of SDR performance with respect to the compute budget.
Here, we use RTX A5000 GPUs for training TF-GridNet and Tesla V100 GPUs for other models\footnote{The equivalent GPU hours using Tesla V100 for TF-GridNet should be even larger since RTX A5000 has faster computing speed.}.
It is shown that BSRNN initially has very high compute efficiency, corresponding to the low model complexity region in Fig.~\ref{fig:scaling} (a).
In contrast, TF-GridNet's high performance comes at a high cost, requiring more computing than other architectures even for its smallest configuration (with only 0.5 M parameters).
Although these RNN-based architectures show excellent performance, they are highly inefficient in scaling up due to the limited parallelization.
\emph{This calls for more explorations on efficiently scalable architectures that can be an SE counterpart to transformers in ASR and LLMs.}

Finally in Fig.~\ref{fig:scaling} (c), we show that the usually reported model parameters do not have a clear correlation with the SE performance when comparing different model architectures.
This is attributed to different SE architectures having highly divergent parameter efficiency.
Therefore, we strongly recommend that \emph{SE studies should provide model complexity as an informative indicator instead of solely model parameters}.
We also note that Fig.~\ref{fig:scaling} lacks sufficient explorations in the large parameter region ($\ge$100 M), which we leave for future work.

\section{Conclusion}
\label{sec:conclusion}
In this paper, we have investigated the scalability of SE models in terms of model architectures, model setups (causal or non-causal), model complexity and sizes, compute budget, and dataset sizes, where we consumed $>$25k GPU hours in total.
Our experiments on combined public corpora verified the great potential of two T-F domain dual-path models (BSRNN and TF-GridNet) respectively in both low and high model complexity conditions.
However, these RNN-based architectures are highly inefficient in scaling up due to the limited parallelization.
This suggests that more exploration in the SE architecture is needed to achieve better efficiency and scalability.
In addition, the ``double descent'' phenomenon on VCTK+DEMAND implies that a small corpus is not suitable to benchmark high-capacity SE models.
Our findings in \S~\ref{ssec:exp_causality} further show the necessity of gathering large-scale multi-domain SE corpora to facilitate the scaling of SE models while mitigating mismatches.
We expect the new insights gained from our investigation can inspire more research towards scalable SE models.

\ifinterspeechfinal
\section{Acknowledgment}
The experiments were done using the PSC Bridges2 system via ACCESS allocation CIS210014, supported by National Science Foundation grants \#2138259, \#2138286, \#2138307, \#2137603, and \#2138296.
\fi

\bibliographystyle{IEEEtran}
\bibliography{mybib}

\end{document}